\documentclass[aps,english,10pt,a4,superscriptaddress,onecolumn,nofootinbib]{revtex4}  
\usepackage{amsmath}
\usepackage{amsthm}
\usepackage{amssymb}
\usepackage{amsfonts}
\usepackage{bbm}
\usepackage{epsfig}
\usepackage{times}
\usepackage{babel}
\usepackage{color}
\usepackage{hyperref}
\usepackage{framed}
\usepackage{changes}
\def\R{{\mathbb{R}}}
\def\C{{\mathbb{C}}}
\def\N{{\mathbb{N}}}

\newtheorem{theorem}{Theorem}
\newtheorem{definition}[theorem]{Definition}
\newtheorem{lemma}[theorem]{Lemma}
\newtheorem{example}[theorem]{Example}

\newtheorem{corollary}[theorem]{Corollary}

\begin{document}
\title{A note on ``Hamiltonian for the zeros of the Riemann zeta function''}
\author{Markus P. M\"uller}
\affiliation{Institute for Quantum Optics and Quantum Information, Austrian Academy of Sciences, Boltzmanngasse 3, A-1090 Vienna, Austria}
\affiliation{Department of Applied Mathematics, University of Western Ontario, London, ON N6A 5BY, Canada}
\affiliation{Department of Philosophy and Rotman Institute of Philosophy, University of Western Ontario, London, ON N6A 5BY, Canada}
\affiliation{Perimeter Institute for Theoretical Physics, Waterloo, ON N2L 2Y5, Canada}

\begin{abstract}
This brief note explicates some mathematical details of Phys.\ Rev.\ Lett.\ \textbf{118}, 130201 (2017), by showing how a version of the operator of that paper can be rigorously constructed on a well-defined linear space of functions.
\end{abstract}

\date{September 27, 2017}

\maketitle

\section{A rigorous definition of the operator}
\label{SecMath}
This section gives a rigorous mathematical definition of a version of the operator $\hat H$ of~\cite{BBM}. The construction here differs slightly from that in~\cite{BBM}, and we will explain in what way at the end of this section.

Let $U:=(-1,\infty)\subset\R$ and $U^+:=(0,\infty)=\{u+1|u\in U\}$. Our first goal is to define a natural inverse of the difference operator (a summation operator $\Sigma$) on certain complex functions over $U^+$. We follow the approach of~\cite{MS}. We specialize Definition 2 from~\cite{MS} --- essentially, the property of Definition~\ref{DefAF} is called ``approximately polynomial of degree $\sigma=0$'' in~\cite{MS}.
\begin{definition}
\label{DefAF}
A function $f:U^+\to\C$ will be called \emph{asymptotically flat} if $f(n+x)-f(n)\longrightarrow 0$ as $n\to\infty$ for all $x\in U^+$.
\end{definition}
\begin{example}
\label{ExLog}
The function $f(x)=\log x$ is asymptotically flat.
\end{example}
\begin{example}
\label{ExFlat}
The function $f(x):=x^{-s}\equiv e^{-s\log x}$ with $s\in\C$ is asymptotically flat if and only if ${\rm Re}(s)>-1$.
\end{example}
\proof
This is an elementary calculus exercise. If ${\rm Re}(s)>-1$, then
\begin{eqnarray*}
   \left|(n+x)^{-s}-n^{-s}\right|&=&\left|-s\int_n^{n+x} t^{-s-1}\, dt\right|\leq |s|\int_n^{n+x}|t^{-s-1}|\, dt=|s|\int_n^{n+x}t^{-{\rm Re}(s)-1}\, dt
   \leq |s|\, |x|\,\max_{t\in [n,n+x]} t^{-{\rm Re}(s)-1}\\
   &=&|s|x n^{-{\rm Re}(s)-1}\quad \stackrel{n\to\infty}\longrightarrow 0
\end{eqnarray*}
since $x>0$ and $-{\rm Re}(s)-1<0$. On the other hand, suppose that ${\rm Re}(s)\leq -1$, then $y\mapsto y^{-{\rm Re}(s)}$ is convex, hence
\[
   |(n+x)^{-s}-n^{-s}|\geq |(n+x)^{-s}|-|n^{-s}|=(n+x)^{-{\rm Re}(s)}-n^{-{\rm Re}(s)}\geq -{\rm Re}(s)n^{-{\rm Re}(s)-1}x\geq x
\]
for $n\geq 1$, and thus the left-hand side is bounded away from zero.
\qed

\begin{definition}
\label{DefSummable}
An asymptotically flat function $f:U^+\to\C$ will be called \emph{summable} if the limit
\begin{equation}
   \lim_{n\to\infty} \left(x\, f(n)+\sum_{\nu=1}^n \left(\strut f(\nu)-f(\nu+x)\right)\right)
   \label{eqLimit}
\end{equation}
exists for every $x\in U$. In this case, the limit will be our definition of $\sum_{\nu=1}^x f(\nu)$, which is a function from $U$ to $\C$.
\end{definition}
For the motivation of this definition see~\cite{MS}.  It is easy to see that asymptotic flatness guarantees that the limit~(\ref{eqLimit}) exists and agrees with the standard finite sum $\sum_{\nu=1}^x f(\nu)$ for all $x\in\N$. If $f$ is summable, then Definition~\ref{DefSummable} extends the definition of $\sum_{\nu=1}^x f(\nu)$ to non-integer values of $x$.
\begin{example}
Consider the asymptotically flat function $f(x)=\log x$ of Example~\ref{ExLog}. We have
\[
   \lim_{n\to\infty}\left(x\log n+\sum_{\nu=1}^n \left(\strut \log(\nu)-\log(\nu+x)\right)\right)=\log\lim_{n\to\infty} \left( n^x \prod_{\nu=1}^n \frac \nu {\nu+x}\right)
   =\log\Gamma(x+1)
\]
for all $x\in U$ (see~\cite[6.1.2]{AS}). Thus $f=\log$ is summable, and $\sum_{\nu=1}^x \log \nu=\log \Gamma(x+1)$ for all $x\in U=(-1,\infty)$.
\end{example}
Definition~\ref{DefSummable} allows us to introduce a linear operator $\Sigma$, acting via $(\Sigma f)(x):=\sum_{\nu=1}^x f(\nu)$. Its domain of definition are the summable functions $f:U^+\to\C$, and every such function is mapped to a complex function on $U$. We can say more about the image of $\Sigma$ --- for example, every function $\Sigma f$, for $f$ asymptotically flat and summable, is ``approximately polynomial of degree $\sigma=1$'' according to the terminology in~\cite{MS}, but we will not need any of this in the following.

We now set $(\Delta f)(x):=f(x)-f(x-1)$ and $(xf)(x):= x\cdot f(x)$, and we introduce the linear operator
\[
   X:=\Sigma x \Delta.
\]
Its domain of definition is
\[
   \mathcal{D}(X):=\left\{
      f:U\to\C\,\,|\,\, x\Delta f:U^+\to\C \mbox{ is asymptotically flat and summable}
   \right\}.
\]
We also consider the operator
\[
   p:=-i\frac d {dx}
\]
on the domain of definition
\[
   \mathcal{D}(p):=\{f:U\to\C\,\,|\,\, f\mbox{ is continuously differentiable}\}.
\]
Analogously, we define the operator $p^+:=-i\,d/dx$, but on the domain of continuously differentiable functions $f:U^+\to\C$. The following calculations will also hold if these domains are chosen differently, e.g.\ by demanding smoothness instead of continuous differentiability. We commit to the choices above for concreteness. We define the operator
\[
   R:=Xp+pX
\]
with domain of definition
\[
   \mathcal{D}(R):=\{f\in\mathcal{D}(X)\cap\mathcal{D}(p)\,\,|\,\, pf\in\mathcal{D}(X)\mbox{ and }Xf\in\mathcal{D}(p)\}.
\]
We are interested in the eigenfunction equation for $R$. As a preparation, we need the following simple statements.
\begin{lemma}
\label{LemDeltaSum}
Let $f:U^+\to\C$ be asymptotically flat and summable. Then $\Delta\Sigma f=f$.
\end{lemma}
\proof
This follows from computing explicitly the difference $\sum_{\nu=1}^x f(\nu)-\sum_{\nu=1}^{x-1}f(\nu)$, using asymptotic flatness and the fact that both defining limits exist.
\qed

\begin{lemma}
\label{LemDeltaEigenf}
Suppose that $\lambda\in\C$ and $f\in\mathcal{D}(R)$ such that $Rf=\lambda f$. Then $(\Delta f)(x)=\alpha\cdot x^{-s}$ for some $\alpha,s\in\C$, where $\Delta f:U^+\to\C$.
\end{lemma}
\proof
Both $Rf$ and $\lambda f$ are functions from $U$ to $\C$. We can apply the difference operator to both sides (yielding functions from $U^+$ to $\C$), and obtain $\Delta R f=\lambda\Delta f$. For every $f\in\mathcal{D}(X)$, we have
\[
   \Delta X f = \Delta \Sigma x \Delta f = x\Delta  f
\]
using Lemma~\ref{LemDeltaSum}, since $x\Delta f:U^+\to\C$ is asymptotically flat and summable. Note also that $\Delta p f = p^+ \Delta f$ if $f\in\mathcal{D}(p)$. Thus, for every $f\in\mathcal{D}(R)$, we have
\[
   \Delta R f=\Delta X \underbrace{pf}_{\in\mathcal{D}(X)}+\Delta p \underbrace{X f}_{\in\mathcal{D}(p)} = x\Delta p \underbrace{f}_{\in\mathcal{D}(p)}+p^+\Delta X \underbrace{f}_{\in\mathcal{D}(X)} = xp^+\Delta f+p^+ x\Delta f.
\]
Solving the corresponding differential equation $(xp^++p^+x)\Delta f=\lambda \Delta f$ proves that $\Delta f(x)=\alpha\cdot x^{-s}$ for some $\alpha,s\in\C$.
\qed
\begin{lemma}
For every $s\in\C$ with ${\rm Re}(s)>-1$, the function $x\mapsto x^{-s}$ is asymptotically flat and summable. Consequently, we can define
\[
   x^{[-s]}:=\sum_{\nu=1}^x \nu^{-s}
\]
for $x\in U$ and ${\rm Re}(s)>-1$, and we have
\[
   x^{[-s]}=\left\{
      \begin{array}{cl}
      	  \zeta(s)-\zeta(s,x+1) & \mbox{if }s\neq 1\\
      	  \gamma+\Psi(x+1) & \mbox{if }s=1,
      \end{array}
   \right.
\]
where $\zeta(\cdot)$ is the Riemann zeta function, $\zeta(\cdot,\cdot)$ is the Hurwitz zeta function, $\gamma$ is the Euler-Mascheroni constant, and $\Psi$ is the digamma function.
\end{lemma}
\proof
Asymptotic flatness has been established in Example~\ref{ExFlat}, and summability and the identities for $x^{[-s]}$ have been proven in~\cite{MS}.
\qed

Using well-known properties of the Hurwitz zeta function, we obtain the following corollary:
\begin{corollary}
$x^{[-s]}$ is continuously differentiable on $U$ for every $s\in\C$ with ${\rm Re}(s)>-1$, and
\[
   \frac d {dx} x^{[-s]} = -s\, x^{[-s-1]}+s\zeta(1+s).
\]
\end{corollary}
\begin{lemma}
\label{LemAlmost}
If $f\in\mathcal{D}(R)\setminus\{\mathbf{0}\}$ satisfies $Rf=\lambda f$ for some $\lambda\in\C$, then either $\lambda=0$, or $f(x)=\alpha x^{[-s]}+\beta$, where ${\rm Re}(s)> 0$ and $\alpha\in\C\setminus\{0\},\beta\in\C$ are suitable constants.
\end{lemma}
\proof
Let $h\in\mathcal{D}(R)$ be any function with $\Delta h=0$, then
\[
   Rh=\Sigma x \Delta p \underbrace{h}_{\in\mathcal{D}(p)} + p\Sigma x\underbrace{\Delta h}_0=\Sigma x p^+ \Delta h=0.
\]
According to Lemma~\ref{LemDeltaEigenf}, we have $(\Delta f)(x)=\alpha x^{-s}$ for some $\alpha,s\in\C$. First consider the case $\alpha=0$, i.e.\ $(\Delta f)(x)=0$, then $Rf=0$, hence $\lambda=0$. Now consider the case $\alpha\neq 0$. Since $f\in\mathcal{D}(R)$, it follows that $f\in\mathcal{D}(X)$, hence $x\Delta f=\alpha x^{-(s-1)}$ is asymptotically flat. According to Example~\ref{ExFlat}, it follows that ${\rm Re}(s)>0$. Consequently, $x^{[-s]}\in\mathcal{D}(R)$.

Let $g(x):=\alpha x^{[-s]}-f(x)$, then $(\Delta g)(x)=0$ and $g\in\mathcal{D}(R)$. A straightforward calculation yields
\begin{equation}
   Rf(x)=R\left(\alpha x^{[-s]}-g(x)\right)=\alpha R x^{[-s]}= i\alpha (2s-1)x^{[-s]}-i\alpha(s-1)\zeta(s)\qquad(s\neq 1),
   \label{eqRImage}
\end{equation}
with $(s-1)\zeta(s)$ replaced by $1$ if $s=1$. For this to equal $\lambda f$, the function $f$ must be a linear combination of $x^{[-s]}$ and a constant function.
\qed

Now we are ready to state the main result:
\begin{theorem}
\label{TheMain}
The set of all $\lambda\in\C$ such that there exists some $f\in\mathcal{D}(R)\setminus\{\mathbf{0}\}$ with $f(0)=0$ and $Rf=\lambda f$ is exactly
\[
   \{0\}\cup\{i(2 s_n-1)\,\,|\,\, s_n\mbox{ is a nontrivial zero of the Riemann zeta function}\}.
\]
This is a subset of $\R$ if and only if the Riemann hypothesis is true.
\end{theorem}
\proof
According to Lemma~\ref{LemAlmost}, $\lambda=0$ is a potential eigenvalue. To see that it is really an eigenvalue, consider the function $\sin(2\pi x)$ on $U$. This function is in $\mathcal{D}(R)$, and $R$ maps it to the zero function since $\Delta\sin(2\pi x)=0$. Since $0^{[-s]}=0$, all eigenfunctions to non-zero eigenvalues must be multiples of $x^{[-s]}$ with ${\rm Re}(s)>0$ due to Lemma~\ref{LemAlmost}. Direct calculation yields
\[
   Rx^{[-s]}=i(2s-1)x^{[-s]}-i(s-1)\zeta(s)\qquad (s\neq 1)
\]
resp.\ $(s-1)\zeta(s)$ replaced by $1$ if $s=1$. Hence $i(2s-1)$ is an eigenvalue if and only if $\zeta(s)=0$. This eigenvalue is real if and only if ${\rm Re}(s)=\frac 1 2$.
\qed

One motivation to consider the operator $R$ originates in an attempt to make the formal calculations of Section 8 in~\cite{MS} rigorous. Studying the properties of the $\Sigma$ operator seems crucial for this. The operator $X$ appears there naturally, since it maps the polynomial $\Sigma x^n$ to the polynomial $\Sigma x^{n+1}$ (note that the domain of definition of $\Sigma$ is larger in~\cite{MS} than in this note). It is then a natural question how this operator interacts with differentiation, which suggests to consider $R$. Since the functions $\Sigma f$ take the value zero at $x=0$, the boundary condition $f(0)=0$ seems quite natural in that context. It can be replaced by $f(-1/2)=0$ without changing the result, since $(-1/2)^{[-s]}=(2-2^s)\zeta(s)$ for $s\neq 1$, cf.~\cite{MS}. It is thus also possible to choose the boundary condition $f(0)=f(-1/2)$, or $(\Delta_{1/2}f)(0)=0$, with $\Delta_{1/2}f(x)=f(x)-f(x-\frac 1 2)$.

The eigenfunctions $x^{[-s]}$ appearing in Theorem~\ref{TheMain} are obviously not elements of $L^2(0,\infty$), since $x^{[-s]}=\sum_{\nu=1}^x\nu^{-s}\sim x^{1-s}/(1-s)$ for $x\to\infty$ if ${\rm Re}(s)< 1$, see e.g.~\cite[eq.~25.11.43]{DLMF}. It is tempting to attempt to introduce an inner product, different from the usual $L^2$ structure, that is based on the sublinear growth of those functions for $0<{\rm Re}(s)<1$. Ignoring many obvious mathematical details, we might, for example, tentatively set $\langle f,g\rangle:=\int_0^\infty \overline{(\Delta_{1/2}f)(x)}(\Delta_{1/2}g)(x)dx$, for functions with $f(0)=f(-1/2)=0$, which might render $p$ formally symmetric under this choice of boundary condition. Conceivable eigenfunctions $x^{[-s]}$ of $R$ with ${\rm Re}(s)>\frac 1 2$ would have finite norm with respect to this inner product. Whether constructions like this can yield more information on the properties of $R$ is unclear.

The operator $R$ in this note differs from the operator $\hat H$ in~\cite{BBM} in the way that the inverse of $\Delta$ is defined. By choosing a suitable space of functions, we enforce that $\Delta f=g$ determines $f$ up to an additive constant. This constant is here fixed by demanding that $f(0)=0$ (since $\sum_1^0 g=0$; for more details see~\cite[Lemma 6]{MS}), and in~\cite{BBM} by demanding that $f(x)\to 0$ as $x\to\infty$. The latter version has the advantage that it can formally be written as a similarity transform of $xp+px$, leading to the further formal calculations in~\cite{BBM}.

\section{Conclusions}
We have given a simple rigorous construction of a version (denoted $R$) of the  operator $\hat H$ of~\cite{BBM} on a suitable vector space of functions. According to Theorem~\ref{TheMain}, its eigenvalues are real if and only if the Riemann hypothesis is true. However, the above does not give us any additional information about the spectrum of $R$, and we have mostly omitted the discussion of how one might equip the linear space with the structure of a Banach or Hilbert space.

By introducing an operator of the form conjectured by Berry and Keating, the construction in~\cite{BBM} aims to open up a new perspective on an existing approach (due to Hilbert and P\'olya). Of course, any given new approach to a hard and extensively studied problem has a very low probability to lead to a  direct solution (some popular accounts may not have emphasized this to the degree we would have preferred). But arguably, this makes the task of finding new perspectives more important, not less so. Studying further properties of the operators $R$ and $\hat H$ therefore seems like an interesting avenue to pursue.

\textbf{Note added.} During the preparation of this note, a comment on Ref.~\cite{BBM} has appeared in Ref.~\cite{Bellissard}. It is not the purpose of this present note to provide an answer to the comment. Instead, for the latter, interested readers are referred to Ref.~\cite{comment2}.

\section*{Acknowledgments}
This research was undertaken, in part, thanks to funding from the Canada Research Chairs program, and it was supported in part by Perimeter Institute for Theoretical Physics. Research at Perimeter Institute is supported by the Government of Canada through the Department of Innovation, Science and Economic Development Canada and by the Province of Ontario through the Ministry of Research, Innovation and Science.

$\strut$\\
See also the references in~\cite{BBM}.

\end{document}